\documentclass[letterpaper,twocolumn,10pt]{article}
\usepackage{usenix,epsfig}
\usepackage{comment}
\usepackage{prettyref}
\newrefformat{sec}{Section~\ref{#1}}
\newrefformat{tab}{Table~\ref{#1}}
\newrefformat{fig}{Figure~\ref{#1}}
\newrefformat{cha}{Chapter~\ref{#1}}
\newrefformat{app}{Appendix~\ref{#1}}
\newrefformat{eqa}{Equation~\ref{#1}}

\usepackage{booktabs}
\usepackage[hyphens]{url}

\usepackage{xspace}
\makeatletter
\newcommand{\ie}{i.e.\@\xspace}

\makeatother

\usepackage{xifthen}
\usepackage[usenames,dvipsnames]{xcolor}
\newcounter{mn}
\newcommand{\superscript}[1]{\ensuremath{{}^{\textrm{\scriptsize #1}}}}

\newcommand{\mntext}[1]{\colorbox{Maroon}{\begin{color}{white}#1\end{color}}}
\newcommand{\mn}[2][]{{\tiny\superscript{\mntext{\arabic{mn}}}}\marginpar{\scriptsize{
  \ifthenelse{\isempty{#1}}
  {\mntext{\parbox{0.95\marginparwidth}{\superscript{\arabic{mn}}~\raggedright{#2}}}}
  {\mntext{\parbox{0.95\marginparwidth}{\superscript{\arabic{mn}}#1 says~:~\raggedright{#2}}}}
}}\stepcounter{mn}}

\begin{document}

\date{}

\title{\Large \bf Censorship Resistance: Let a Thousand Flowers Bloom?}

\author{
{\rm Tariq Elahi}\\
University of Waterloo\\
{\rm \texttt{tariq.elahi@uwaterloo.ca}}\\
\and
{\rm Steven J. Murdoch}\\
University College London\\
{\rm \texttt{s.murdoch@ucl.ac.uk}}\\
\and
{\rm Ian Goldberg}\\
University of Waterloo\\
{\rm \texttt{iang@cs.uwaterloo.ca}}\\
} %

\maketitle

\subsection*{Abstract}

This paper
argues that one of the most important decisions in designing and deploying censorship resistance systems is whether one set of
system options should be selected (the best), or whether there should
be several sets of good ones. We model the problem of choosing these options
as a cat-and-mouse game and show that the best strategy depends on
the value the censor associates with total system censorship versus
partial, and the tolerance of false positives. If the censor has a low
tolerance to false positives then choosing one censorship resistance
system is best. Otherwise choosing several systems is the better choice,
but the way traffic should be distributed over the systems depends on
the tolerance of the censor to false negatives. We demonstrate that
establishing the censor's utility function is critical to discovering the
best strategy for censorship resistance.

\section{Achieving censorship resistance}

Anonymous communication systems provide a powerful tool for Internet
denizens to speak out against oppression by protecting their
identities. Censorious regimes try to limit access to such systems to
further control their populations. 
This paper discusses 
how to best provide censorship-resistant access to these
anonymous communication systems.  
We use Tor as a case study, but the principles discussed here apply
equally well to other systems.  Any such system is built of at least
two components: one to resist blocking by IP address, and another to
resist blocking based on payload. Additionally, the system might be
designed to resist active probing among other properties as deemed
suitable to combat the threats posed to the system and its users and
operators.

\subsection{IP-address blocking resistance}

IP-address blocking resistance is achieved by
having a wide
variety of IP addresses that will provide access to the network,
possibly in combination with distributing IP addresses to users of the
network such that the censor is not able to discover them.
``Bridges''~\cite{tor-blocking} are Tor's approach to this part of the
problem.  Here, Tor users in countries that do not block access to
the Tor network are encouraged to run a Tor node that is not listed
in the public Tor directory.  IP addresses of bridges are distributed
such that an adversary with limited resources (in particular, IP
addresses and Gmail email addresses) is unable to enumerate all
bridges.  By making it as easy as possible to set up bridges, it was
hoped there would be many, and the distribution strategy was designed
such that Tor users in countries that block Tor will be able to find
at least one IP address that the censor does not know about.  Tor's strategy has been successful;
while China was able to discover and block a large proportion of
bridges,  
Tor users in other countries from which access to the Tor
network is blocked by IP address still can use bridges.

An alternative approach to be proposed is to have a relatively small
number of bridge nodes accessible via a very large number of IP
addresses.  These IP addresses are shared with other services. This is
done to avoid wasting IP addresses and, more importantly, also to
force censors to block access to important services that are not
banned in the country. This threat of collateral damage, through false
positives, is to discourage attempts to block the IP addresses
associated with the bridges.
These goals are achieved through ``Decoy
Routing''~\cite{usenix11-telex,ccs2011-cirripede}, where network
traffic destined to the anonymous communication system is
steganographically tagged and sent to an IP address in a network which
supports decoy routing.  The border router for this network, or a
specialized computer designed for this task, detects the tag and
routes the traffic to the anonymous communication system.

Another similar approach is ``triangle
routing''~\cite{triangleboy}.
In itself, this does not provide any additional IP addresses for
accessing the anonymous communication network, so bridges or a similar
approach is still required.  What triangle routing achieves is to
permit these bridges to be on low-bandwidth connections yet still
offer high performance.  Network traffic exiting the censored
country does get relayed via the bridge node, to a network-entry node,
but return traffic is sent directly from the network-entry node to the
user, spoofing the IP address of the bridge.

\subsection{Payload-blocking resistance}

The advantage for the censor
of IP-address based blocking is that standard IP routers are, by
definition, capable of routing traffic based on IP address and can
thus redirect or block traffic destined for certain IP addresses.
However, more sophisticated routers and specialized censorship
equipment is capable of looking within packets (known as Deep Packet
Inspection, DPI) and blocking network traffic which fulfils particular
criteria. Therefore censorship resistance systems must also disguise
packet content. For the purpose of this paper the payload-blocking
resistance will also include resisting blocking based on port number,
packet timing and packet size and other payload characterisitics.

\subsubsection{Impersonating nothing}

One approach to payload-blocking resistance is for network traffic to
have no static characteristics.  Payload data is encrypted to appear
indistinguishable from random; packet sizes and timings are
also randomized.  The encryption necessitates some sort of key
agreement.
This could be unauthenticated, such as simply sending the
key in the clear at the start of the communication, or performing
ephemeral Diffie-Hellman.  The former option is vulnerable to
network blocking equipment capable of extracting the key and
decrypting subsequent traffic.  The latter is resistant to passive
attack, but vulnerable to an active man-in-the-middle.  Alternatively,
key agreement could be performed out of band or the client could be authenticated based on
credentials exchanged out of band.

One system that takes the ``impersonating nothing'' approach is
obfs3~\cite{obfs3}.
This is a wrapper around the Tor bridge protocol,
which works by sending the obfuscation key in the clear at the start
of the communication.  Subsequent traffic is encrypted by AES under
keys derived from the obfuscation key.  As with normal Tor bridges,
obfs3 can use any TCP port, and the choice made by the obfs3-bridge
must be communicated to the user, along with the IP address.  No
attempt is currently made to hide packet timing or size.  While obfs3
does not perform any authentication or integrity checks, it wraps the
unmodified Tor protocol which does perform both.

Another system with similar goals is Dust~\cite{dust}.  Like obfs3,
network traffic is indistinguishable from random, but it aims to
resist both active and passive attack by relying on a password
exchanged out of band.  Dust also performs integrity checks and provides replay
protection.  While Dust can be used over both UDP and TCP (obfs3 is
TCP only), it does not provide a reliable in-order transport if sent over UDP,
so an
additional layer would be needed before a TCP-based protocol such as
Tor could be used with the UDP variant. Like obfs3, Dust also does not itself hide packet timings or
lengths.

The advantage of the ``impersonating nothing'' approach is that it has
no static payload signature that could be programmed into DPI
equipment.  However, such traffic is also unlike almost anything else
seen on the Internet, so if DPI equipment can detect it, the false
positive rate would likely be very low.  One such test would be to
measure the entropy of a communication stream, using one or more of the
many tests for random number generators.  Any traffic with a value
that is higher than expected would be blocked.  Performing such a test
would be challenging because it cannot be expressed as a regular
expression, which is the common interface exposed by DPI equipment
for configuring new blocking rules.
Additionally, some entropy tests
have high RAM and CPU usage and so would be infeasible to run directly
on high-bandwidth DPI equipment which only has a handful of CPU cycles
for each packet, and store a few tens of bytes for each stream.
Therefore a staged approach would be needed: efficient initial tests
either for entropy or to exclude known protocols would be performed on
all traffic, and only selected packets would be sent for subsequent
processing.

Another way of blocking ``impersonating nothing'' protocols is through
a whitelist.  Only protocols which match a particular (perhaps
port-dependent) signature would be permitted.  As even encrypted
traffic is commonly surrounded by an unencrypted header, it would be
possible to find a set of DPI rules which would permit a substantial
portion of network traffic.  However, protocols which were not
explicitly permitted would be blocked, and so there could be a
substantial false-positive rate, especially for more obscure
protocols.  This false-positive rate would increase if more protocols
became indistinguishable from random, so one way for protocol
designers to help censorship resistance would be to hide any protocol
characteristics, even if they have no need for censorship resistance
themselves.

\subsubsection{Impersonating something}

As an alternative, the censorship-resistance scheme could impersonate
a particular network protocol.  Tor already does this to some extent,
by using TLS for its outermost cryptography layer.  Initial versions
of Tor made no attempt to appear like web browsing, and so Tor TLS
connections included a number of distinctive characteristics such as
static fields in certificates and an unusual set of ciphersuites.
Later, the Tor TLS options were made closer to that of common web
browsers and web servers, by randomizing certificate fields and
selecting ciphersuites commonly seen on the Internet.

However, Tor differs from typical encrypted web browsing in one
important way, which is that connections between Tor relays depends on bidirectional
authentication, rather than only server-to-client.
In the initial
version of Tor, the client certificate was sent unencrypted and thus
could be used to block traffic.  In later versions of Tor, the client
certificate is sent during a renegotiation phase, which is encrypted.
Unfortunately, the fact that renegotiation is being performed can be
inferred from the plaintext in the network traffic,
because the type of a TLS record is not
encrypted, and a client certificate is of a different type from the
application data which would be expected after TLS key exchange.

Efforts are underway to make Tor traffic even closer to encrypted web
browsing.  So far these include using more commonly seen
Diffie-Hellman parameters, and extending the expiry time of
certificates to be more plausible. 
The next step to be taken will be to disguise the
renegotiation step, by implementing client-to-server authentication
within the Tor protocol itself.

However, impersonating TLS is not a silver bullet.  TLS is very
common, and blocking TLS would cut off access to many useful Internet
services, but countries have been willing to do so.  Iran, in
particular, has blocked TLS across much of the country for periods of
time~\cite{iran_partial}.
While partial, these blocks were at precisely the time that access to
an anonymous communication network would be most important.  For this
reason, Tor supports ``pluggable transports'', which are external
programs responsible for obfuscating Tor traffic.  obfs3 is one
such pluggable transport, but it was always intended that there be
many such available, taking a variety of approaches.  Such approaches
do not only include the obfuscation technique, but also development
practices---while Tor is open source, there is no reason that an
pluggable transport could not be distributed as an obfuscated binary if
that was considered to make it hard to reverse engineer and
block.
Existing pluggable transports include impersonating HTTP traffic like
StegaTorus~\cite{stegotorus} and impersonating Skype traffic like
SkypeMorph~\cite{skypemorph}.

\subsection{Scanning resistance}

Being accessible at a wide variety of IP addresses, and disguising
payload, are sufficient for resisting passive blocking.  However, more
sophisticated adversaries may also perform active attacks by scanning
IP addresses and detecting whether they are entry points for an
anonymous communication network.

One option would be to proactively
scan IP addresses---perhaps identified through some survelliance means
to reduce the address space---
to check if anonymous communication software is
listening; another is to target scanning based on network
surveillance.  China has taken the latter approach, by recording which IP
addresses are contacted by computers in China over TLS where the
ciphersuite list matches the one used by Tor~\cite{china-blocking}.
Shortly after such a connection, another computer in China probes the
IP address and attempts to establish a Tor circuit, and if successful,
the IP address is blocked.  This approach has allowed China to almost
completely block access to non-obfuscated Tor bridges.

To resist such probing, there needs to be a way for anonymous
communication nodes to distinguish between legitimate access and
probing, and either fail to respond to probing or return content which
the censor considers innocuous.  This goal can be achieved by sharing
a secret between the legitimate user and the access node,
and designing
an authentication scheme for which an authentication failure is
indistinguishable from the node not being an access node.  One such
scheme is BridgeSPA~\cite{wpes11-bridgespa}, which encodes an authentication
token into the TCP initial sequence number and timestamp field, and
simply rejects connections for which the authentication check fails.

\section{Putting it all together}
For each of the components of a censorship resistance system, there
are a wide variety of options available.
Each of the options available has its own advantages and there is a trade-off between
in terms of
overhead, implementation and deployment difficulty, and security.  One
common question, however, is whether to put limited development effort
into making one censorship resistance scheme that is highly resistant
to blocking, or to spread effort over multiple, less robust methods.
Which option is the best is more of a question of economic incentives
rather than a purely technical decision.  As such it depends on how
both the censor and developer of the censorship resistance system
value particular situations.

\subsection{Censor costs}

The costs of the censor are mainly in terms of political capital and
financial capital.  Political capital is spent by false positives (by
annoying users of services that are not intended to be blocked, \ie
collateral damage) and false negatives (by
failing to block sites that they should, \ie information
leakage). Financial capital is spent on blocking equipment and
engineering time adding and testing new blocking rules and the
variable cost of deploying such a system to suit the scale of traffic
being monitored.

Neither cost function is necessarily in direct proportion to the
underlying quantities; there could easily be discontinuities.
It may also be that these
costs are step functions where costs stay the same until some threshold
limit after which the costs skyrocket. Knowing these threshold limits
is key in deploying, and overcoming, economically sound censorship
systems.

\subsection{Modelling the Cost of Censorship}
We now move towards establishing a model for quantifying the cost of
censorship. As noted above it is difficult to establish the true total cost
of censorship, such as the cost of developing and training the
technology, maintaining it, and the associated cost of
failure -- political and financial. To get around this shortcoming we
shift our focus to the accuracy of the censor's tools
and use it as a proxy to understand the magnitude of the costs.

\subsubsection{Technological Limits}
Censorship technology is limited by shortcomings in the languages used
to define classification expressions, the computational and memory
costs of real-time processing and the partial view of the attack
surface amongst other considerations.
It is important, then, to take in to account the rate at
which objects of interest are misclassified. The two types of
errors---false positives and false negatives---govern the confidence
the censor has in their censorship apparatus. The prevalence of each of
these type of errors provides an important input for both the censor
and the defender in defining their respective strategies. 

The base rate bias further thwarts accurate
detection in favour of misclassification when the censor's false
positive rate is comparable in size to, or larger than, the incidence rate of
the defender's traffic.
In such a case, a significant proportion, if not almost all, of the
traffic selected for censoring will be false positives.

\subsubsection{False Positives}
From the censor's perspective, false positives are the legitimate
traffic, and users, that were misclassified and blocked---the
\emph{collateral damage}. The censor naturally seeks to keep this as
low as is acceptable. Without censor cooperation it is difficult to
learn the cost of collateral damage, if indeed the censor is even able
to evaluate this cost itself.  
However, we can
assign a proxy value based on how many citizens were potentially
inconvenienced. While this most likely does not align with the
censor's values they can provide an independent measure of the cost of
collateral damage in terms of disgruntled citizens and hence upward
pressure on political costs.

As a strategy, collateral damage has been leveraged by numerous
censorship resistance systems. However, in most cases the defender
assumes an all-or-nothing approach to censorship, which can be limiting
when the censor is content with partial blocking. The defender's, and
client's, costs are in terms of effort to impersonate legitimate
protocols since this would give the highest rate of false positives.
The defender must also be careful to keep the rate of incidence of
defender traffic low if they seek to
maximize the amount of collateral damage caused.

\subsubsection{False Negatives}
The censor tries to prevent as many clients as it can from circumventing its
blocks---termed \emph{information leakage}. Due to the limits of technology
it is unable to identify all of them. The cost to the censor is
entirely in terms of political capital. Since it is hard to quantify
the censor's cost, indeed he may also not be able to evaluate it in
measurable terms, we use circumventing client connections as an
objective proxy. Again, the censor may not have the same evaluation;
nevertheless the fact remains that the censorship apparatus has failed
in some manner and the circumventing connections are points of success
for the defender.

The defender's aim is always to have as much, if not all, of its
traffic classified as a (false) negative. The use of traffic morphing,
steganography, and encryption to name a few techniques, are
instrumental in achieving this goal.

\subsubsection{Bringing it Together}

The cost to the censor will depend on numerous factors, including the
equipment and expertise necessary to implement censorship, but for
simplicity we will assume that the utility to the censor is only a
function of the false positives and false negatives of the blocking.
Moreover, we will assume that the cost of blocking a protocol is
proportional to how popular the protocol is, whereas in practice users
are more likely to tolerate some protocols being blocked than others.
However the analysis techniques we outline do not depend on these
assumptions and could be generalised if necessary.

As an example, we shall use the following 
function to model the censor's utility in our analysis. The constants $C$ and $D$
control the tolerance of the censor to false positives and false
negatives respectively. The variable $t$ is the percentage of the
target protocol blocked (i.e. the true positives) and $f$ is the
percentage of other traffic blocked (i.e. the false positives).

\begin{equation}
U_c = 100 - 200 (1 - e^{C (\frac{100 - t}{D} + f)})
\label{eqa:uc}
\end{equation}

This function allows a wide range of plausible censor utility
functions to be modelled, and results in a value between $-100$ (maximum
dissatisfaction) and $+100$ (maximum satisfaction)

\section{The Censorship Game}

We model the cat-and-mouse game of censor vs. censorship resistant
communications provider (the \textit{distributor} in our terminology)
as a multi-round
game. The distributor moves first, and provides
software which impersonates one or more protocols and distributes user
traffic over these protocols according to some probability
distribution. The censor then obtains a copy of the software, and is
able to establish which protocols are being impersonated and in which
proportion. The impersonation is sufficiently good that the censor
must choose to either block a protocol entirely -- blocking both cover
traffic (causing false positives) and the distributor's traffic
(causing true positives), or leaving it entirely unblocked.

We let the censor move second because it is likely that the censor can
move faster than the distributor because the distributor must roll out
new software to many thousands of users in order to change strategy
whereas the censor needs only to make a configuration change.
In each round the censor will choose a blocking strategy to maximize
their utility function. The goal of the traffic distributor is to find
the traffic probability distribution function such that the censor's
best strategy is the one the distributor finds most acceptable. This
will be the equilibrium strategy since if either party changes their
choice, they will decrease their own utility.

An interesting consequence of this model is that the utility function
of the censorship resistant communications provider does not matter,
as all they can do is choose between the collection of scenarios which
the censor has decided to be optimum for a particular strategy of the
distributor. Therefore, as long as the distributor's utility function
is monotonically decreasing in terms of the true positive rate, the
same equilibrium will be reached regardless of the function's shape.

\subsection{False-positive intolerant censor}

We first consider a censor with low tolerance to false positives. We
define this to mean that there is at least one protocol which they are
unwilling to block (a \emph{critical} protocol), even if this would
result in blocking 100\% of the distributor's traffic. In this case
the distributor should choose critical protocol and send all
censorship-resistance traffic over it. The censor will not block it,
and so 100\% of traffic will get through.  Any alternative strategy
for the distributor would be less good, as choosing multiple critical
protocols would be more effort for no gain, and choosing a
non-critical protocol for some traffic might lead the censor to block
it.

\subsection{False-positive tolerant censor}

A more interesting case is where there is no such critical protocol.
To give a concrete example, assume that the distributor can
impersonate all of the top 6 protocols from a 2014 survey of US
Internet traffic~\cite{sandvine-Internet-phenom}: 
YouTube (13.25\%), HTTP (8.47\%), BitTorrent (5.03\%), SSL (2.63\%), MPEG (2.44\%), and Amazon Video (2.37\%).

\begin{figure}
	\includegraphics[width=\columnwidth]{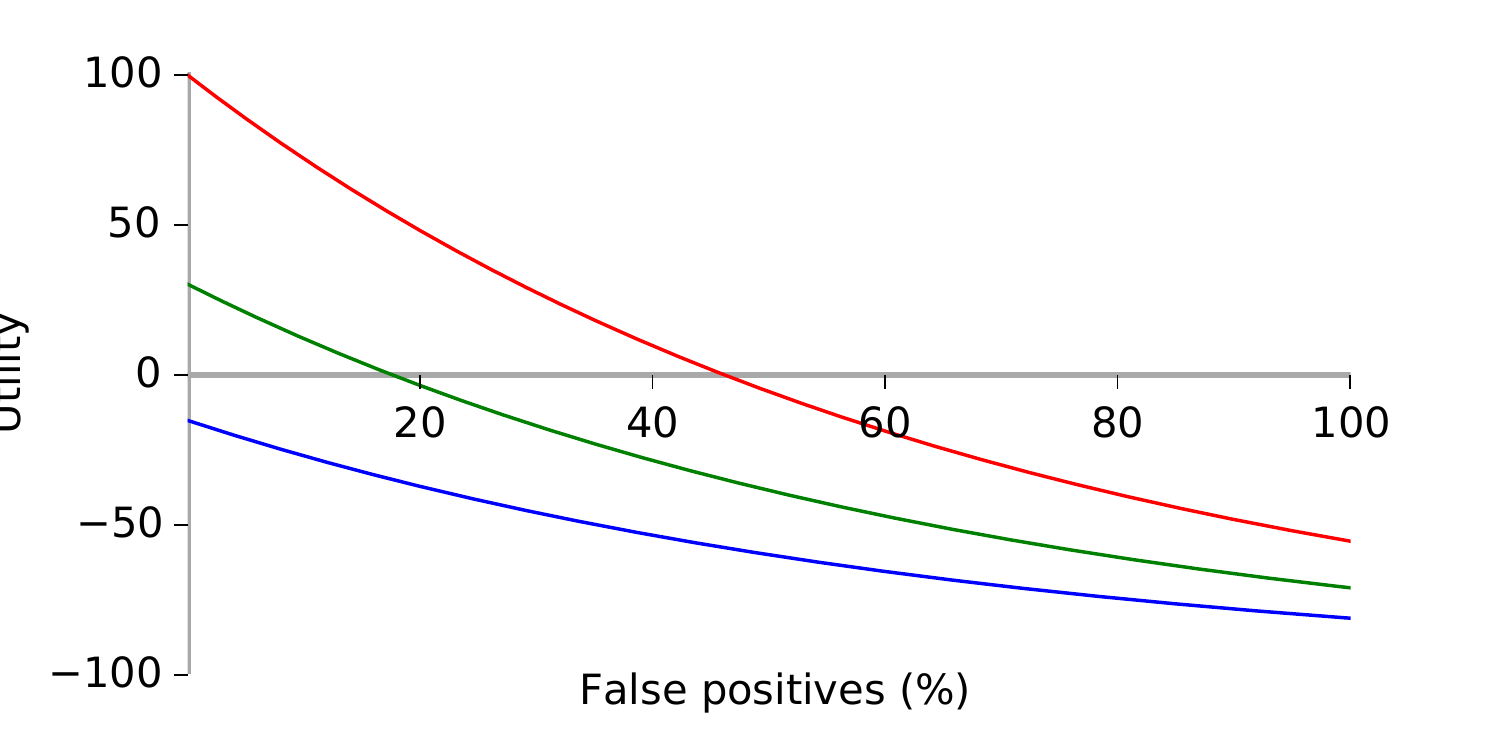}
	\caption{\label{fig:censor1}Utility of a censor with
		high false-positive and false-negative tolerance.}
\end{figure}

As the censor utility function, we use \prettyref{eqa:uc} with $C=-0.015$ and
$D=1.75$. This is illustrated on \prettyref{fig:censor1} for three values of
true positive rates: 100\% (top), 50\% (middle) and 0\% (bottom).

We now need to compute the censor utility function for all
combinations of censor strategy and distributor strategy. The censor
can choose to block any selection of protocols of the 6 considered
(there is no reason to block any others). As a result there are $2^6 =
64$ scenarios.

The distributor can choose any probability distribution, but we
exclude any distribution where the traffic distributed over protocol
$a$ is greater than the proportion distributed over protocol $b$ when
the proportion of cover traffic going over protocol $b$ is greater
than that of $a$. We do this because if any excluded scenario were
chosen, if $a$ and $b$ were swapped, the censor utility function would
be lower for every censor scenario (assuming the censor prefers a
lower false-positive rate).

\begin{figure}
	\includegraphics[width=\columnwidth]{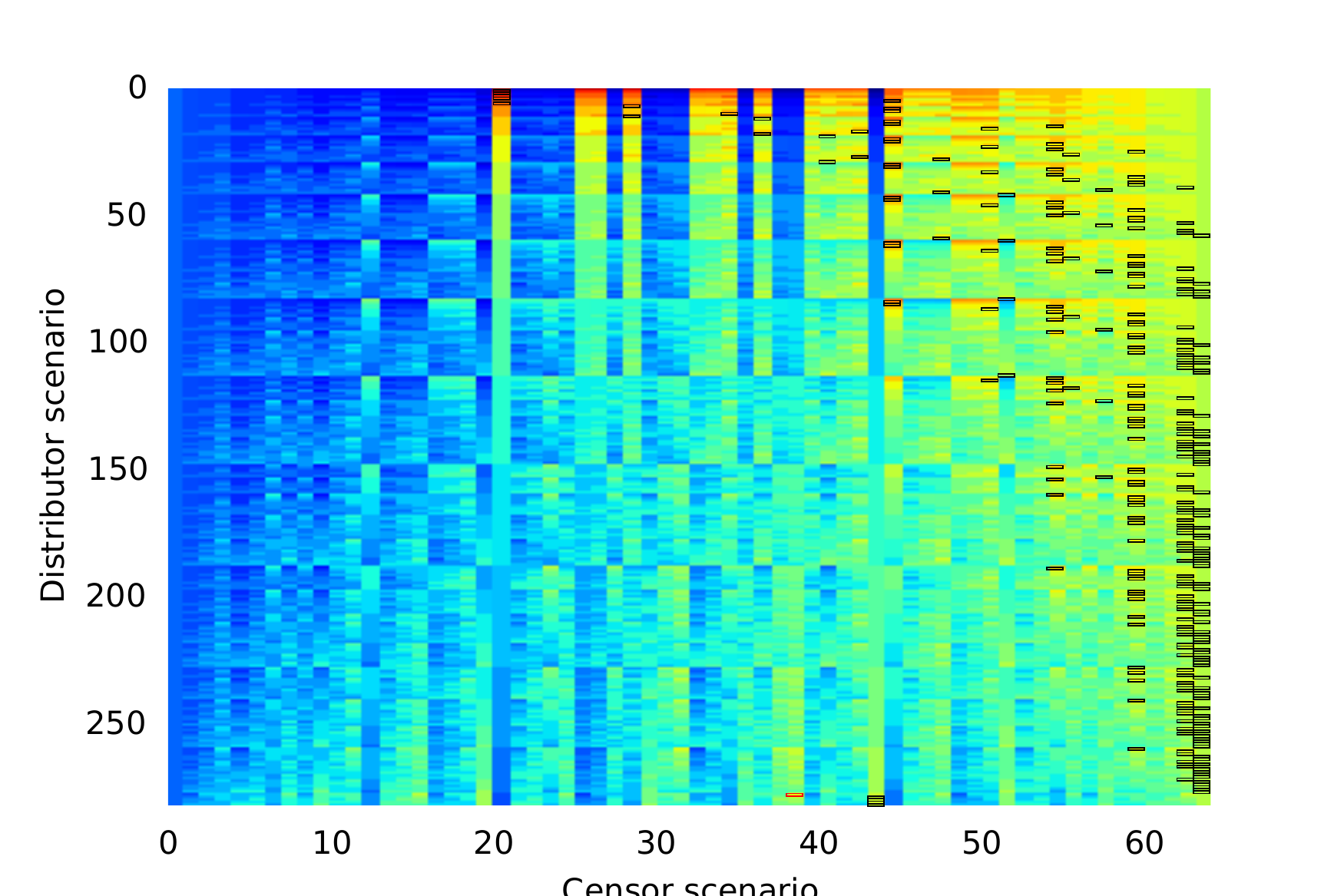}
	\caption{\label{fig:heatmap1}Utility heatmap for a censor with high false-positive and
		false-negative tolerance. The cell with the red outline is the best strategy for the censor and the equilibrium.} \end{figure}

Even making this assumption there are still an infinite number of
distributor scenarios, so to reduce the number we quantize all
proportions to be a multiple of 5, resulting in 282 distributor
scenarios.
The result of simulating all scenarios is shown on
\prettyref{fig:heatmap1}, where blue is low utility and red is high utility.
The censor scenarios to the left have low false positives; those to the right
have high false positives. The distributor scenarios at the top have traffic
heavily skewed to the protocols with most cover traffic; those at the bottom
have traffic more evenly distributed over protocols. Rectangles show the
optimum censor strategy for each distributor strategy (red for the equilibrium
and black for others).

Even small changes in the distributor scenarios results in large
changes in optimum censor scenario, but the equilibrium is to
distribute traffic quite evenly over protocols, but not completely.
The top 5 protocols should each get 20\% of traffic with the
6th not used at all. The censor will block protocols 2, 3, 4 and 5
allowing 20\% of the distributor traffic through. Were the attacker
to block protocol 1, the additional false-positives
would not justify the extra 20\% of true positives. Were the
distributor to move some traffic onto protocol 6, it would be blocked
because it has a smaller false-positive cost.

\subsection{Another false-positive tolerant censor}

\begin{figure}[ht]
	\includegraphics[width=\columnwidth]{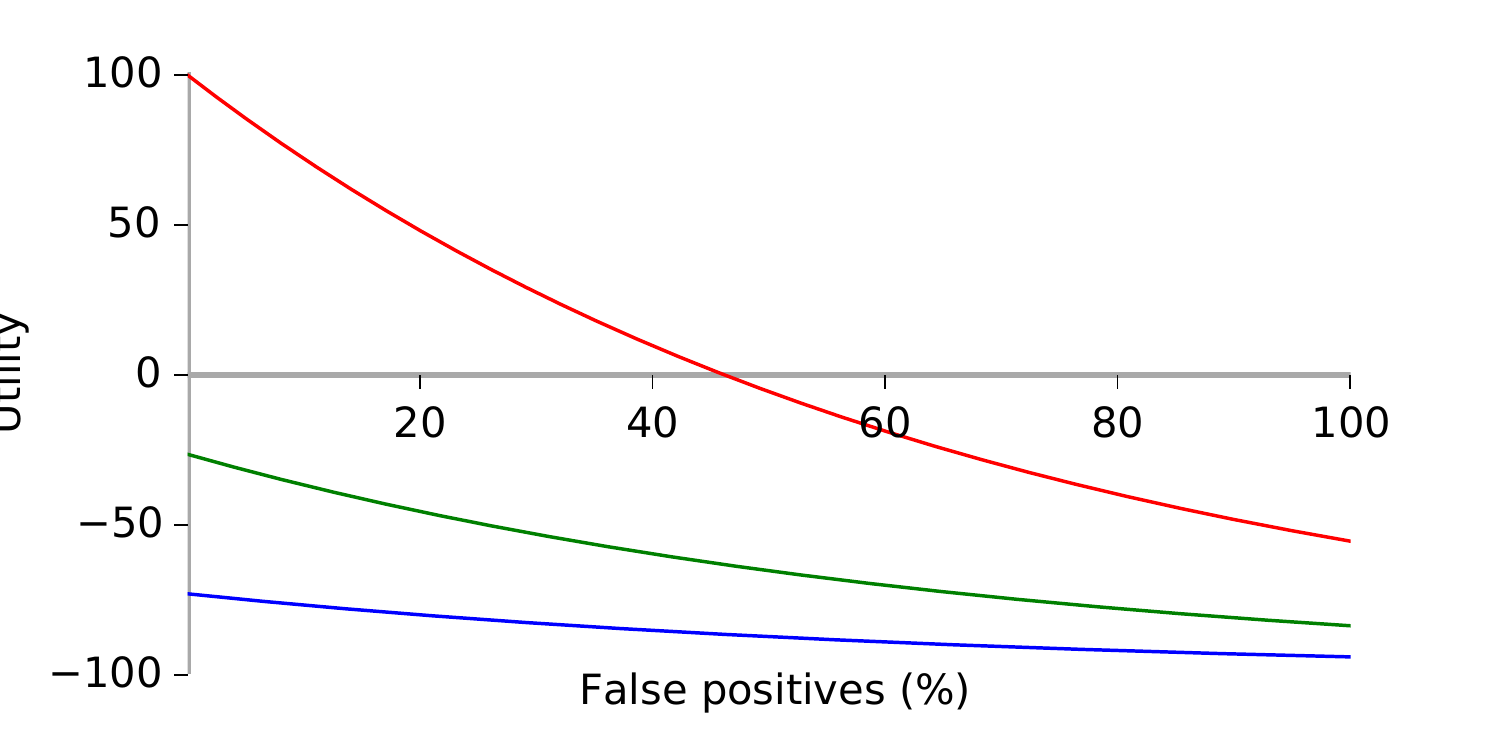}
	\caption{\label{fig:censor2}Utility for a
		censor with high false-positive and low false-negative
		tolerance.}
\end{figure}

\begin{figure}[ht]
	\includegraphics[width=\columnwidth]{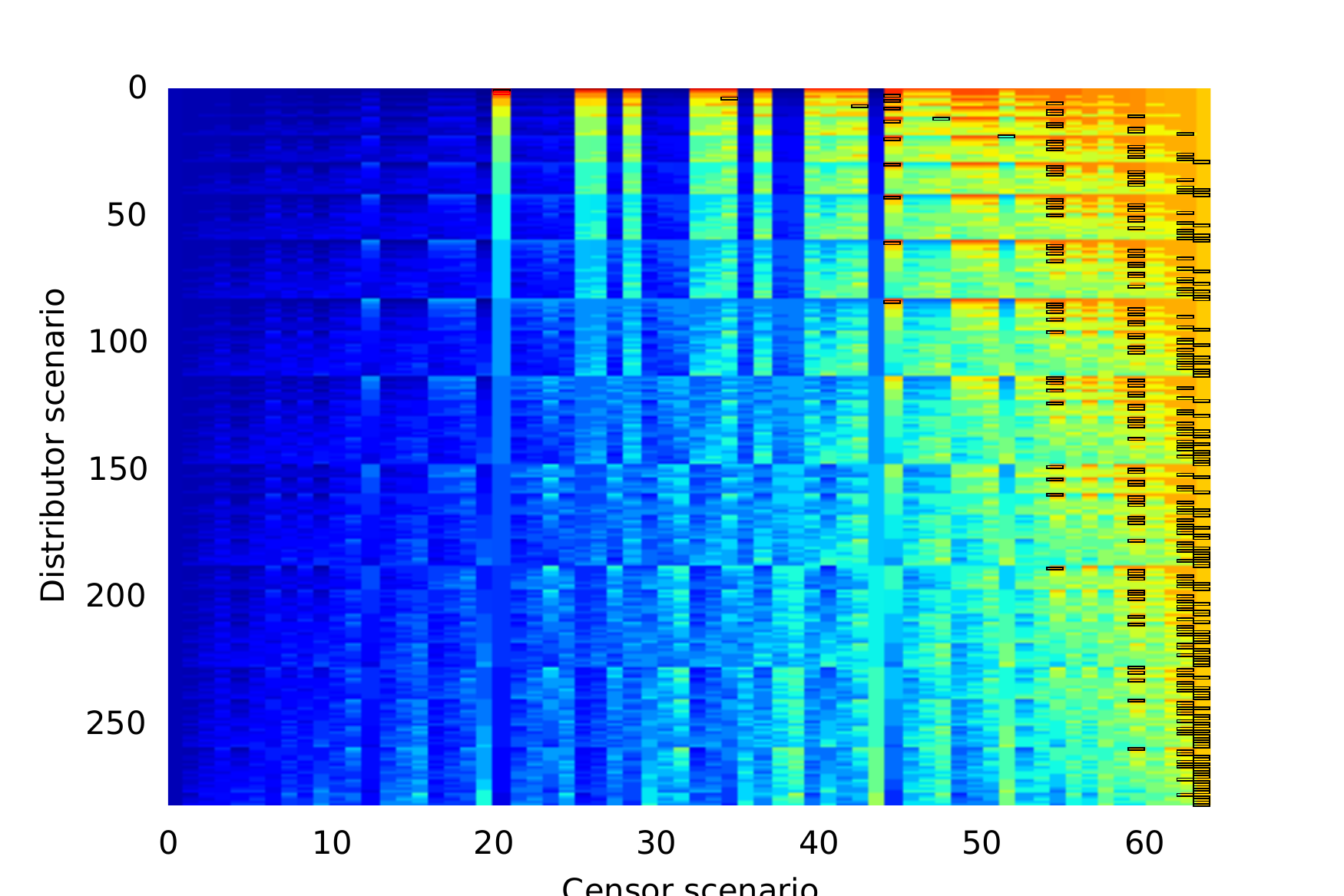}
	\caption{\label{fig:heatmap2}Utility for a censor with high
false-positive and low false-negative tolerance. The cell with the red outline is the best strategy for the censor and the equilibrium.} \end{figure}

Let us now consider a censor who is equally tolerant to
false-positives, but far less tolerant to false-negatives than before,
by changing $D$ from $1.75$ to $0.75$ with the result shown
in \prettyref{fig:censor2}
Now a 50\% false
negative rate is significantly below zero for all false positives,
whereas before it would have been positive if the false positive rate
was low. The resulting simulation is significantly different, as can
be seen in \prettyref{fig:heatmap2}.

Now the optimum strategy for the censor is almost always to block many
protocols, resulting a high false-positives (the right hand side of
the graph). The equilibrium strategy is for 95\% of traffic to be
distributed on protocol 1 and 5\% to be distributed on protocol 2. The
censor will block protocol 1, but leave protocol 2 unblocked. This
lets only 5\% of distributor traffic through, but it is better than
the 0\% which almost every other strategy results in. For example,
sending 100\% over protocol 1 results in protocol 1 being blocked.
Sending 80\% over protocol 1 and 10\% over protocol 2 results in both
protocols being blocked. Putting only 5\% over protocol 2 is small
enough that the extra benefit to the censor of blocking it is not
large enough to justify the high false positives.

\section{Conclusions}

We have shown how to discover the equilibrium strategy for a censor-distributor cat-and-mouse
 game and show how the censorship
resistance provider should distribute traffic over different protocols
to impersonate. The equilibrium depends on the censor's utility
function, but not that of the censorship resistance provider. If the
censorship resistance provider is in the happy position of being able
to impersonate a protocol which the censor is not willing to block,
then only one protocol should be impersonated. If however the censor
is willing to block even the most popular protocol, the best strategy
is depends on the detail of the censor's utility function, and will
result in distributing traffic over some, but perhaps not all
protocols. Discovering this utility function is difficult, but if
approximations can be found it may be possible to improve the
approaches of current censorship resistance systems.

{\footnotesize \bibliographystyle{acm}
\bibliography{censorship-games}}

\end{document}